\documentclass[conference]{IEEEtran}

\usepackage{amsmath,amssymb,amsfonts}
\usepackage{dcolumn,tabularx,booktabs}
\usepackage{multirow}
\usepackage{makecell}
\usepackage[backend=biber,sorting=none]{biblatex}
\usepackage{algorithmic}
\usepackage{graphicx}
\usepackage{textcomp}
\usepackage{tikz}
\usepackage{pgfplots}
\usepackage{pgfplotstable}
\usepackage{xcolor}
\usepackage{svg}
\usepackage{siunitx}
\usetikzlibrary{fit}
\usetikzlibrary{patterns}
\usepackage{xspace}
\usepackage{wheelchart}
\usepackage{subcaption}
\usepackage{siunitx}
\usetikzlibrary{positioning,fit,calc}
\usepgfplotslibrary{groupplots}

\newcommand*\circled[1]{\tikz[baseline=(char.base)]{
            \node[shape=circle,draw,inner sep=1pt] (char) {#1};}}

\pgfplotsset{compat=1.18}
\def\BibTeX{{\rm B\kern-.05em{\sc i\kern-.025em b}\kern-.08em
    T\kern-.1667em\lower.7ex\hbox{E}\kern-.125emX}}

\definecolor{KITgreen}     {RGB}{0,150,130}
\definecolor{KITblue}      {RGB}{70,100,170}
\definecolor{KITblack}     {RGB}{0,0,0}
\definecolor{KITgray}      {gray}{0.3}
\definecolor{KITlightgray} {gray}{0.84}
\definecolor{KITyellow}    {RGB}{252,229,0}
\definecolor{KITorange}    {RGB}{223,155,27}
\definecolor{KITlightgreen}{RGB}{140,182,60}
\definecolor{KITred}       {RGB}{162,34,35}
\definecolor{KITpurple}    {RGB}{163,16,124}
\definecolor{KITbrown}     {RGB}{167,130,46}
\definecolor{KITcyan}      {RGB}{35,161,224}

\makeatletter
\def\ps@IEEEtitlepagestyle{%
  \def\@oddfoot{\mycopyrightnotice}%
  \def\@evenfoot{}%
}
\def\mycopyrightnotice{%
  {\footnotesize
    \parbox{\textwidth}{
      \centering
      \copyright~2025 IEEE. Personal use of this material is permitted. Permission from IEEE must be obtained for all other uses, in any current or future media, including reprinting/republishing this material for advertising or promotional purposes, creating new collective works, for resale or redistribution to servers or lists, or reuse of any copyrighted component of this work in other works by sending a request to \texttt{pubs-permissions@ieee.org}.
    }
  }%
  \gdef\mycopyrightnotice{}
}
\makeatother

\begin{document}

\title{Real-Time Graph-based Point Cloud Networks on FPGAs via Stall-Free Deep Pipelining}

\newcommand{\blind}{false}

\DeclareRobustCommand*{\IEEEauthorrefmark}[1]{%
  \raisebox{0pt}[0pt][0pt]{\textsuperscript{\footnotesize #1}}%
}

\ifthenelse{\equal{\blind}{false}}
{
    \author{
    \IEEEauthorblockN{
    Marc Neu\IEEEauthorrefmark{*1},
    Isabel Haide\IEEEauthorrefmark{+},
    Timo Justinger\IEEEauthorrefmark{*},
    Till Rädler\IEEEauthorrefmark{*}, 
    Valdrin Dajaku\IEEEauthorrefmark{*},
    Torben Ferber\IEEEauthorrefmark{+} and
    Jürgen Becker\IEEEauthorrefmark{*}}
    \IEEEauthorblockN{\IEEEauthorrefmark{*}Institute for Information Processing Technology,
    Karlsruhe Institute of Technology, Germany}
    \IEEEauthorblockN{\IEEEauthorrefmark{+}Institute of Experimental Particle Physics,
    Karlsruhe Institute of Technology, Germany}
    \IEEEauthorblockA{
    \IEEEauthorrefmark{1}marc.neu@kit.edu}}
}
{}


\maketitle


\begin{abstract}
 Graph-based Point Cloud Networks (PCNs) are powerful tools for processing sparse sensor data with irregular geometries, as found in high-energy physics detectors. However, deploying models in such environments remains challenging due to stringent real-time requirements for both latency, and throughput. In this work, we present a deeply pipelined dataflow architecture for executing graph-based PCNs on FPGAs. Our method supports efficient processing of dynamic, sparse point clouds while meeting hard real-time constraints. We introduce specialized processing elements for core graph operations, such as GraVNet convolution and condensation point clustering, and demonstrate our design on the AMD~Versal~VCK190. Compared to a GPU baseline, our FPGA implementation achieves up to 5.25× speedup in throughput while maintaining latencies below  $\mathbf{10~\text{\bfseries\si{\micro s}}}$, satisfying the demands of real-time trigger systems in particle physics experiments. An open-source reference implementation is provided.
\end{abstract}

\begin{IEEEkeywords}
Point Cloud Network, Graphs, FPGA, Real-Time, High-Performance-Computing, Dataflow, Latency
\end{IEEEkeywords}

\section{Introduction}
\begin{figure}
    \centering
    \includegraphics[width=\linewidth]{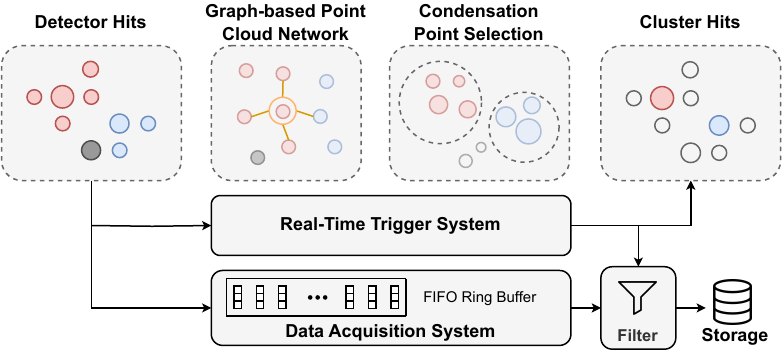}
    \caption{Simplified application example for graph-based PCNs: Machine Learning-based clustering  in high energy physics.}
    \label{fig:trg:application}
\end{figure}

Point clouds are a natural choice for representing heterogeneous, sparse sensor data, well suited as input for machine learning algorithms~\cite{guo2021, bello2020}.
Point Cloud Networks (PCNs) have demonstrated strong performance in classification and semantic segmentation tasks involving irregular geometries~\cite{qi2016,wang2018}.
Graph-based PCNs, which use message passing over intermediate graphs to aggregate features between points, are particularly effective for such data~\cite{wei2023, wemmer2023}. Instead of relying on fixed output structures or predefined object counts, these models learn to associate inputs with dynamically determined object representations~\cite{qu2020,iiyama2020}.
Real-world applications for such models are high energy physics experiments~\cite{abe2010a, CMS:2006myw}, where detector geometries are highly irregular.
In these experiments large amounts of raw data is processed by an online filtering stage, as depicted in Figure~\ref{fig:trg:application}.

However, these systems employ stringent system requirements~\cite{Lai:2025gac,CMS:2000mvk} making a deployment of sparse, graph-based PCNs especially challenging:\\
(1) Online filtering decisions must be made within $\SIrange{1}{10}{\micro\second}$, due to limited buffer capacity.
(2) Detector snapshots (events) are sampled at rates of up to 40 million events per second (EPS), imposing high throughput requirements.
(3) Technical constraints enforce strict in-order processing, requiring hard real-time guarantees for all algorithms.

These requirements introduce a set of challenges for deploying sparse, graph-based PCNs in real-time systems:\\
(1) Processing dynamic input sparsity in the range of \SIrange{5}{20}{\percent} and heterogeneous workload efficiently
(2) Real-Time Dynamic Graph Building and Message Passing on FPGAs
(3) Real-Time Clustering through a suitable implementation of the condensation point selection algorithm~\cite{kieseler2020}.

To close the gap in the current literature, this work explores how to deploy graph-based PCNs on FPGAs under real-time constraints, ensuring high throughput and low latency.

We identify our contributions in this work as follow:\\
(1) We propose a deployment approach for graph-based PCNs tailored for real-time FPGA execution. Our method is compatible with state-of-the-art data-flow compilation toolchains, such as FINN~\cite{blott2018} and hls4ml~\cite{hls4ml}.
(2) We demonstrate our approach in an end-to-end implementation, comparing its inference performance against a GPU baseline. Our results show that our FPGA implementation achieves higher throughput while meeting the strict real-time requirements.
(3) We release a standalone, open-source implementation of our architecture, providing a reference for future work in real-time PCN deployment.

\section{A Motivating Example}
\begin{figure}
    \centering
    \includegraphics[width=\linewidth]{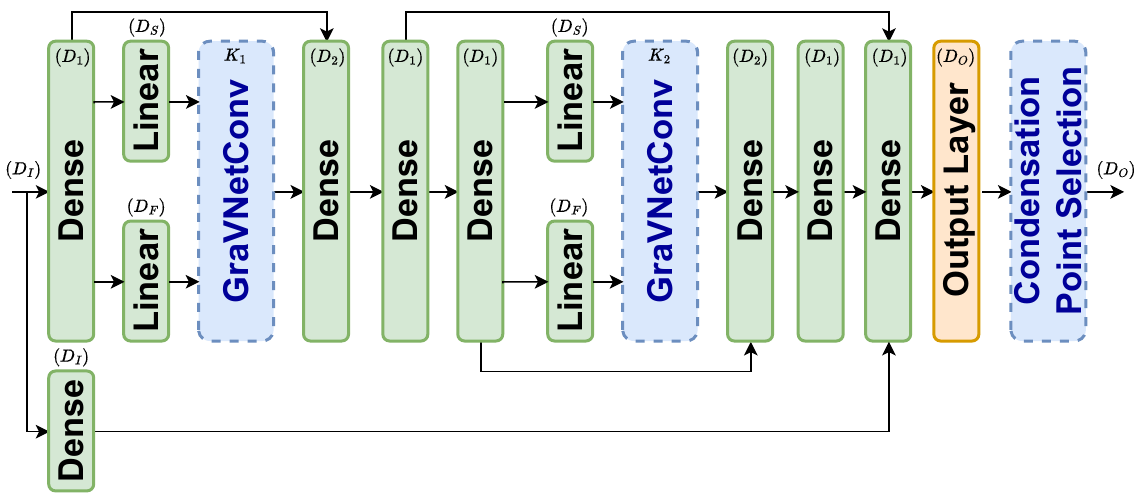}
    \caption{We implement the following neural network based on~\cite{qasim2019, wemmer2023} with dimensions $D_I=5$, $D_1=16$, $D_2=32$, $D_S=6$, $D_F=8$, $D_O=9$, and hyperparameter $K=8$. For trainable layers, we choose a weight sparsity of \SI{40}{\percent}.}
    \label{fig:statement:network}
\end{figure}

To highlight the challenges of deploying graph-based PCNs in large scientific experiments, we present a real-world case study. Specifically, we consider the clustering of energy depositions in the electromagnetic calorimeter (ECL) at Belle~II~\cite{abe2010a,kou2018}. 
Due to the scintillation decay time of the CsI(Tl) crystals and the response of the shaper electronics, sampling \qty{8}{\mega EPS} is sufficient for this system.

We implement the network from Figure~\ref{fig:statement:network}, consisting of dense layers, linear layers without activation function, GraVNetConv layers~\cite{qasim2019}, and the condensation point clustering layer~\cite{kieseler2020}.

\begin{figure}
    \pgfdeclareplotmark{pytorch-tensorrt+fp16}
    {\pgfuseplotmark{diamond}}
    \centering
    \small
    \begin{tikzpicture}
        \begin{axis}[
            set layers=axis on top,
            width=\columnwidth,
            height=6cm,
            clip mode=individual,
            xlabel={Events per Second (\si{EPS})},
            ylabel={Latency (\si{\micro s})},
            xmode=log,
            ymode=log,
            log basis x=10,
            log basis y=10,
            xmin=5e3, xmax=2e7,
            ymin=1, ymax=5e3,
            visualization depends on=value \thisrow{run} \as \runtype]

        \draw[draw=none,fill=KITgreen!30] (8e6,1) rectangle (2e7,10);
        \node[draw,align=center,shape=circle, inner sep=1pt] (upgradelabelconstraints) at  (rel axis cs: 0.8,0.38) {2};
        \draw[->] (upgradelabelconstraints) -- (8e6,10);

        \draw[draw=none,fill=KITgreen!50] (8e6,1) rectangle (2e7,5);
        \node[draw,align=center,shape=circle, inner sep=1pt] (currentlabelconstraints) at  (rel axis cs: 0.78,0.15) {1};
        \draw[->] (currentlabelconstraints) -- (8e6,2);

        \addplot[
            only marks,
            mark=\runtype,
            mark size=3pt,
            restrict expr to domain={\thisrow{point_num}}{64:64},
            restrict expr to domain={\thisrow{pareto}}{0:0},
            KITgray!30!white
        ] table [
            x=eps, 
            y=latency_us, 
            col sep=comma
        ] {figures/pareto_gpu.csv};


        \addplot[
            only marks,
            mark=\runtype,
            mark size=3pt,
            restrict expr to domain={\thisrow{point_num}}{256:256},
            restrict expr to domain={\thisrow{pareto}}{0:0},
            KITgray!30!white
        ] table [
            x=eps, 
            y=latency_us, 
            col sep=comma
        ] {figures/pareto_gpu.csv};

        \addplot[
            const plot mark right,
            mark=\runtype,
            thick,
            mark size=3pt,
            restrict expr to domain={\thisrow{point_num}}{32:32},
            restrict expr to domain={\thisrow{pareto}}{1:1},
            unbounded coords=discard,
            KITcyan
        ] table [
            x=eps, 
            y=latency_us, 
            col sep=comma
        ] {figures/pareto_gpu.csv};

        \addplot[
            const plot mark right,
            mark=\runtype,
            thick,
            mark size=3pt,
            restrict expr to domain={\thisrow{point_num}}{64:64},
            restrict expr to domain={\thisrow{pareto}}{1:1},
            unbounded coords=discard,
            KITpurple
        ] table [
            x=eps, 
            y=latency_us, 
            col sep=comma
        ] {figures/pareto_gpu.csv};

        \addplot[
            const plot mark right,
            thick,
            mark size=3pt,
            mark=\runtype,
            restrict expr to domain={\thisrow{point_num}}{128:128},
            restrict expr to domain={\thisrow{pareto}}{1:1},
            unbounded coords=discard,
            KITbrown
        ] table [
            x=eps, 
            y=latency_us, 
            col sep=comma
        ] {figures/pareto_gpu.csv};

        \addplot[
            const plot mark right,
            mark=\runtype,
            thick,
            mark size=3pt,
            restrict expr to domain={\thisrow{point_num}}{256:256},
            restrict expr to domain={\thisrow{pareto}}{1:1},
            unbounded coords=discard,
            KITred
        ] table [
            x=eps, 
            y=latency_us, 
            col sep=comma
        ] {figures/pareto_gpu.csv};

        \end{axis}

        \node[] (legendsizetitle) at  (rel axis cs: -0.85,0.275) {\scriptsize Event Size};
        \node[fill=KITcyan,draw,minimum height=0.2cm,minimum width=0.2cm,label=0:\scriptsize 32,] (legendsize1) at  (rel axis cs: -0.9,0.2) {};
        \node[fill=KITpurple,draw,minimum height=0.2cm,minimum width=0.2cm,label=0:\scriptsize 64,] (legendsize2) at  (rel axis cs: -0.9,0.125) {};
        \node[fill=KITbrown,draw,minimum height=0.2cm,minimum width=0.2cm,label=0:\scriptsize 128,] (legendsize3) at  (rel axis cs: -0.7,0.2) {};
        \node[fill=KITred,draw,minimum height=0.2cm,minimum width=0.2cm,label={[name=legendsize4label]0:\scriptsize 256},] (legendsize4) at  (rel axis cs: -0.7,0.125) {};
        \node[fit=(legendsizetitle)(legendsize1)(legendsize2)(legendsize3)(legendsize4)(legendsize4label), draw] {};
        
    \end{tikzpicture}
    \caption{Pareto-front for the neural network shown in Figure~\ref{fig:statement:network}. The batch size is varied from $\{1,...,4096\}$.}
    \label{fig:gpu:pareto}
\end{figure}
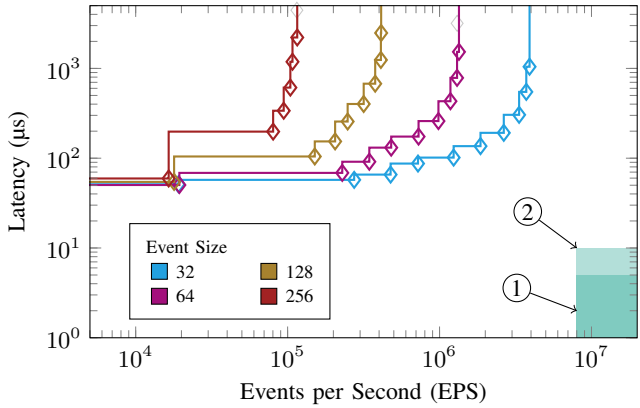

As a baseline we deploy the neural network on a NVIDIA~GPU~L40S with BF16 precision and report latency and throughput in Figure~\ref{fig:gpu:pareto}. The model is compiled from PyTorch~\cite{pytorch} using Tensor-RTs~\cite{tensorrt} benchmark tool \textit{trtexec}. We enable all available performance optimization.
We evaluated various state-of-the-art frameworks, including PyTorch~Geometric~\cite{pytorch}, ZenTorch~\cite{zentorch}, and FastGraphCompute~\cite{FGC}, but none yielded better performance.
Two legal design spaces for the ECL are shown in green: \circled{1} for the current system~\cite{Kim:2017uee} and \circled{2} assuming an improved version of the DAQ system~\cite{Aihara:2024zds}. 
This example demonstrates, that state-of-the-art GPUs are not sufficient for our target application and that a custom FPGA-based accelerator architecture is necessary.

\section{Concept}
We leverage deep pipelining to efficiently compute PCN inference in real time by: (1) exploiting input sparsity in variable-sized point clouds, (2) maximizing pipeline utilization through a tailored dataflow architecture, and (3) semi-automatically mapping networks onto this architecture.

\subsection{Sparsity}
A detector consists of $N_{tot}$ sensors, where the data of each unique sensor $p$ is represented by a feature vector $\mathbb{R}^F$. Thus, we define an event as a point cloud $\mathcal{P} = \{ p_0, ... , p_{N-1}\} \subseteq \mathbb{R}^F$ of $N$ points with dimension $F$, where $N \in \mathbb{N}: N \leq N_{tot}$.
The number of points $N$ and therefore the input sparsity for a given event can be adjusted through the sensor sensitivity, i.e. the energy threshold.
As we have full control of the detector, we choose the energy threshold such that we are able to derive an upper bound $\overline{N}$ of simultaneously hit sensors in a given event.
If an event contains more than $\overline{N}$ hits, we drop additional points from the input set.
By our definition we fulfill hard real-time requirements, if our accelerator is capable of processing up to $\overline{N}$ simultaneously hit sensors in a given event.
The actual number of points $N$ for a given event is unknown a-priori, and thus is considered an additional input for the network.

Generally, the sensor data for each event $P$ is represented in a matrix $X_{tot}\in M_{N \times F}(\mathbb{R})$. However, due to the variable number of points per event, this matrix contains a large number of zero rows, i.e. structured sparsity. 
To process such matrices efficiently, either a suitable dynamic sparse tensor operator, or alternatively a compaction mechanism exploiting the sparsity pattern is required. 
We choose the latter option, as the worst-case execution time of dynamic operators is hard to predict.
Thus, for each event $P$, we replace $X_{tot}$ with a compressed matrix $X  \in M_{\overline{N} \times F}$. 
In order to retain the original position of the point in $X_{tot}$, we introduce an additional matrix $Y \in M_{\overline{N} \times 1}$, containing the original position of each non-zero point in the array.

Our accelerator expects the matrices $X,Y$ and the variable $N$ as an input for a given point cloud network, in order to compute the inference result $Z$.

\subsection{Architecture}
We design an architecture consisting of pipelined hardware modules. 
Each module has a number of inputs $N_i$ and a number of outputs $N_o$. The parallelization of a given module is defined through the \texttt{PAR} parameter, which will be explained later on. 
We require that all modules perform computations in order, simplifying latency and throughput analysis. 
In addition, we require that all modules represent single-rate actors, similar to FINN~\cite{blott2018} and hls4ml~\cite{hls4ml}. In order to optimize the deployment for PCNs in general, we define three classes of modules: (1) Point Processing Elements, (2) Graph Processing Elements, (3) Topology Elements.

\textbf{Point Processing Elements (PPEs)} perform independent computations on all points of an event, for example a dense layer. PPEs map operations which are executed point-wise similar to a vector processor. We require $N_i, N_o = 1$, the parallelization may be arbitrarily chosen with $1 < \texttt{PAR} < \overline{N}$. As a result the module has \texttt{PAR} input and output streams. 

\textbf{Graph Processing Elements (GPEs)} perform computations on all points of a given event. Thus $N_i$, $N_j$, and \texttt{PAR} may be chosen freely. GPEs act as an abstraction layer for more complex operators and usually are composed of multiple modules internally. 

\textbf{Topology Elements} do not perform any computation, but facilitate data orchestration between modules. 
Topological dataflow operators such as \texttt{forks} or \texttt{joins} are mapped onto the hardware architecture through these modules.
In addition, changes in hardware parallelization between different stages of a neural network may be facilitated through topological elements, trading resource utilization vs throughput and latency.

In order to avoid performance degradation in all modules, we enforce that (1) all child pipelines must have the same initiation interval, (2) there must be no stalls at any given time inside the processing element and (3) the parallelization must be parameterisable through the \texttt{PAR} parameter. All modules achieve an initiation of $I_{init} = \lceil \frac{\overline{N}}{\texttt{PAR}} \rceil$. 

\begin{figure}
    \centering
    \includegraphics[width=\linewidth]{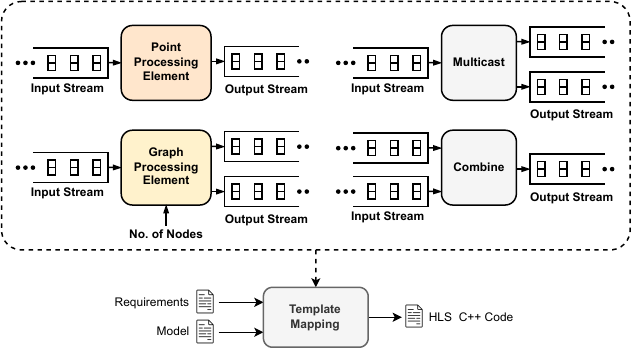}
    \caption{Conceptual view of our deployment approach, showing our proposed actor library used in our dataflow architecture. $\texttt{PAR} = 2$ for all modules.}
    \label{fig:dataflow}
\end{figure}

\subsection{Deployment}
In order to deploy a graph-based PCN onto our dataflow architecture, layers in the network must be mapped onto implementations templates. 
Figure~\ref{fig:dataflow} depicts our deployment flow. 
Similar to FINN~\cite{blott2018}, we use pattern matching on the dataflow graph to replace respective network layers by their processing element. 
In addition, we require that all model-specific layers must be available in our template library. 
Contrary to existing frameworks, our concept is limited to PCNs mappable through our dataflow actors in Figure~\ref{fig:dataflow}. 
However, this allows us to develop PEs running at 100\% pipeline efficiency at all times, maximizing throughput and minimizing latency for a given configuration.

In order to showcase our programming model, we develop three custom PEs for the network presented in Figure~\ref{fig:statement:network}.

(1) We implement a PPE to map batched dense layers for PCN inference. 
The control flow is adapted to repeated computations for every point in an event differentiating the mapping from a regular dense layer defined in FINN~\cite{blott2018}.

\begin{figure}
    \centering
    \includegraphics[width=\linewidth]{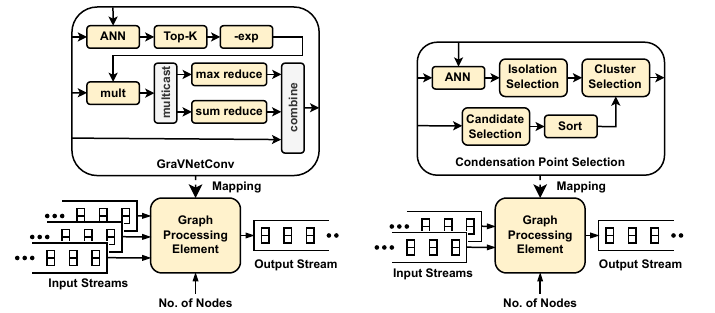}
    \caption{Exemplary mapping of a GraVNetConv layer and condensation point selection layer onto Graph Processing Elements with $PAR = 2$.}
    \label{fig:dataflow:gpe}
\end{figure}

(2) We implement a GPE for the GraVNetConv operator as depicted in Figure~\ref{fig:dataflow:gpe}. These processing elements combines the graph building and message passing steps in one module. Dynamic graph building itself consist of an all-nearest-neighbor (\texttt{ANN}) implementation followed by a hierarchical Top-K sort (\texttt{Top-K}). By implementing an  \texttt{ANN} algorithm, our approach differs significantly from previous static graph building approaches on FPGAs~\cite{neu2024}.
For the message passing step, sorted features are retrieved from BRAM ping-pong buffers and multiplied with exponentially weighted distances (\texttt{exp}, \texttt{mult}). 
Neighborhood aggregation is realized through \texttt{max reduce} and \texttt{sum reduce} operators. 
Finally, output features are aggregated into a single FIFO interface. 

(3) We implement a GPE for the Condensation Point Selection operator as depicted in Figure~\ref{fig:dataflow:gpe}. 
This processing element realizes a clustering algorithm based on a feature space. 
The isolation between cluster seeds are calculated through an \texttt{ANN} followed by an isolation selection. 
The actual cluster points are selected based on a priority value which is sorted (\texttt{sort}), before the final cluster selection step is performed.

\section{Related Work}
PCN acceleration has been explored in ASICs like Point-X~\cite{zhang2021b} and FPGA accelerators like DeepBurning-GL~\cite{liang2020}, but these architectures overlook the specific throughput and latency demands of our application.

Owing to the graph-based aggregation steps, our work is closely aligned with FPGA overlays like FlowGNN~\cite{sarkar2023} and GNNBuilder~\cite{abi-karam2023}. However, unlike our approach, these methods do not account for dynamic graph building.

The GravNetConv layer was explored in prior work~\cite{neu2024}, which implemented only a single layer and fell short of meeting the required throughput due to limited pipelining efficiency.

Our implementation of the condensation point selection algorithm is the first real-time implementation on FPGA.

\section{Evaluation}
\subsection{Experimental Setup}

For demonstration, we implement a dataflow accelerator for the PCN from Figure~\ref{fig:statement:network} on an AMD~Versal~VCK190. 
We use AMD~Vitis~2024.2 for our implementation study and measure the performance metrics \textbf{on the hardware platform} via the XRT runtime. 
For the reference implementation, we use performance metrics shown in Figure~\ref{fig:gpu:pareto}.

We evaluate our system in two different ways: (1) In the end-to-end evaluation, we measure the execution time including the memory transfers $\Delta t_{e2e} = t_{e2e,1}-t_{e2e,0}$ through trace files provided by the XRT Runtime. This evaluation scheme is fairer for the comparison to the GPU baseline. (2) In the compute evaluation, we present cycle-accurate measurements obtained from profiling $\Delta t_c = t_{c,1}- t_{c,0}$, excluding the overhead of AXI transfers. This evaluation scheme is closer to the application field.

In total, we evaluate six different versions, five successfully implemented, of our accelerator with varying parameters as described in Table~\ref{tab:params}. All measurements are repeated 1000 times, we report the \SI{95}{\percent} quantile for all performance metrics both on the GPU and on the FPGA board.
For compute-only measurements $\Delta t_c$ we make sure that all values are identical, verifying deterministic execution times required to fulfill hard real-time constraints.

\begin{table}
    \centering
    \scriptsize
    \caption{Overview of implementation configurations.}
    \label{tab:params}
    \begin{tabular*}{\columnwidth}{@{\extracolsep{\fill}}cccccc}
        \toprule
        \multirow{2}[3]{*}{Version} &  \multicolumn{3}{c}{Parameters} & \multicolumn{2}{c}{Frequency} \\
        \cmidrule(lr){2-4} \cmidrule(lr){5-6} &  Precision & N & \texttt{PAR} & $f_{kernel}$ & $f_{mem}$ \\ \midrule
        A & \SI{8}{bit}   & 32  & 2 & \SI{290}{\mega\hertz} & \SI{312}{\mega\hertz} \\
        B & \SI{16}{bit}  & 32  & 2 & \SI{260}{\mega\hertz} & \SI{312}{\mega\hertz} \\
        C & \SI{8}{bit}   & 64  & 2 & \SI{280}{\mega\hertz} & \SI{312}{\mega\hertz} \\
        D & \SI{16}{bit}  & 64  & 2 & \SI{249}{\mega\hertz} & \SI{312}{\mega\hertz} \\
        E & \SI{8}{bit}   & 128 & 1 & \SI{127}{\mega\hertz} & \SI{312}{\mega\hertz} \\
        F & \SI{16}{bit}  & 128 & 1 & -/- & -/- \\
        \bottomrule
    \end{tabular*}
\end{table}

\subsection{Throughput Measurement}

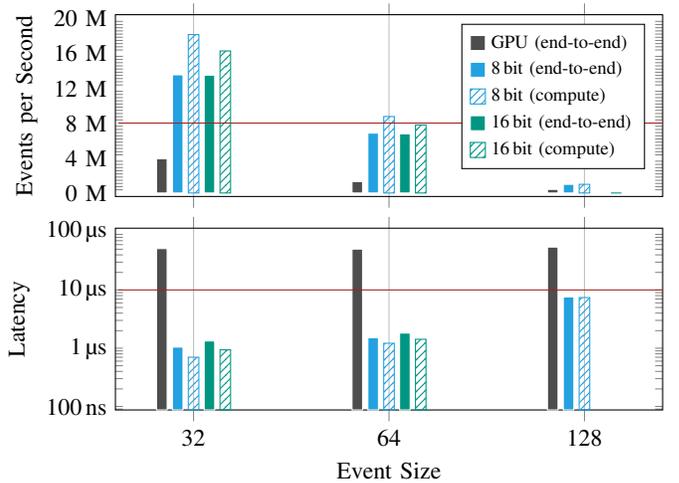
\begin{figure}
    \small
    \centering
    \pgfplotsset{scaled y ticks=false}
    \begin{tikzpicture}
    \begin{groupplot}[
        group style={
            group size=1 by 2,
            x descriptions at=edge bottom,
            vertical sep=12pt,
        },
    ]
    \nextgroupplot[
            ybar,
            width=\columnwidth,
            height=4cm,
            ylabel={Events per Second},
            ymin=0, ymax=2e7,
            xtick={1,2,3},
            ytick={0,4e6,8e6,12e6,16e6,20e6},
            yticklabels={0~M,4~M,8~M,12~M,16~M,20~M},
            legend style={
                font=\scriptsize,
                at={(0.8,0.95)},
                anchor=north, 
                draw=black,
                fill=white,
                legend cell align={left}
            },
            enlarge y limits=0.02,
            enlarge x limits=0.2,
            bar width=4pt,
            major y tick style = transparent,
            xmajorgrids = true,
            minor y tick num=10,
            legend image code/.code={%
                \draw[#1] (0cm,-0.1cm) rectangle (0.2cm,0.1cm);
            }
        ]
            \addplot[style={white,fill=KITgray,mark=none}] coordinates {(1,3913289.728) (2,1305927.0) (3,412736.512)};
            \addlegendentry{GPU~(end-to-end)}
            \addplot[style={white,fill=KITcyan,mark=none}] coordinates {(1,13521610.312868804) (2,6849612.869780431) (3,991185.5629563757)};
            \addlegendentry{\SI{8}{bit}~(end-to-end)}
            \addplot[style={white,KITcyan,pattern color=KITcyan,pattern=north east lines,mark=none}] coordinates {(1,290e6/16) (2,280e6/32) (3,127e6/128)};
            \addlegendentry{\SI{8}{bit}~(compute)}
            \addplot[style={white,fill=KITgreen,mark=none}] coordinates {(1,13480890.278520592) (2,6771423.139553146) (3,0)};
            \addlegendentry{\SI{16}{bit}~(end-to-end)}
            \addplot[style={white,KITgreen,pattern color=KITgreen,pattern=north east lines,mark=none}] coordinates {(1,260e6/16) (2,249e6/32) (3,0)};
            \addlegendentry{\SI{16}{bit}~(compute)}
            \addplot[KITred,sharp plot,update limits=false,shorten >=-10mm,shorten <=-10mm]  coordinates {(1,8e6) (4,8e6)};    
    \nextgroupplot[
            ybar,
            width=\columnwidth,
            height=4cm,
            ylabel={Latency},
            ymin=1e-7, ymax=100e-6,
            xtick={1,2,3},
            yticklabels={0,\SI{100}{\ns},\SI{1}{\micro s},\SI{10}{\micro s},\SI{100}{\micro s}},
            xlabel={Event Size},
            xticklabels={
                {32},
                {64},
                {128}
            },
            ymode=log,
            log origin=infty,
            log basis y=10,
            enlarge y limits=0.02,
            enlarge x limits=0.2,
            bar width=4pt,
            major y tick style = transparent,
            xmajorgrids = true,
            minor y tick num=10,
            legend style={draw=none},
            legend image code/.code={%
                \draw[#1] (0cm,-0.1cm) rectangle (0.2cm,0.1cm);
            }
        ]
            \addplot[style={white,fill=KITgray,mark=none}] coordinates {(1,51.2695e-6) (2,50.293e-6) (3,54.4434e-6)};
            \addplot[style={white,fill=KITcyan,mark=none}] coordinates {(1,1.045e-6) (2,1.514e-6) (3,7.598e-6)};
            \addplot[style={white,KITcyan,pattern color=KITcyan,pattern=north east lines,mark=none}] coordinates {(1,203/290e6) (2,316/260e6) (3,940/127e6)};
            \addplot[style={white,fill=KITgreen,mark=none}] coordinates {(1,1.331e-6) (2,1.81899e-6) (3,0)};
            \addplot[style={white,KITgreen,pattern color=KITgreen,pattern=north east lines,mark=none}] coordinates {(1,244/260e6) (2,354/249e6) (3,0)};
            \addplot[KITred,sharp plot,update limits=false,shorten >=-10mm,shorten <=-10mm]  coordinates {(1,10e-6) (4,10e-6)};        
    \end{groupplot}
    \end{tikzpicture}  
     
    \caption{Measurements for successful implementations from Table~\ref{tab:params}.}
    \label{fig:evaluation}
\end{figure}

We measure the throughput of our accelerator by executing our FPGA kernel with a large batch size of 16384 to reduce influence of the nondeterministic host system. Figure~\ref{fig:evaluation} depicts throughput measurements of our accelerator, design parameters for the accelerator are listed in Table~\ref{tab:params}. The horizontal red line indicates our application requirement of \SI{8}{MEPS}.

We reach the application throughput requirements for event sizes of 32 and 64, but not for 128. This result is expected, as the GraVNetConv complexity is bound by $\mathcal{O}(N^2)$.
Additionally, we observe that for smaller event sizes the difference between end-to-end and compute measurements is larger. For $N = 32$ the architecture is clearly memory bound by the DDR interface on the VCK~190.
In the end-to-end evaluation, we achieve a speedup of \textbf{{3.46}x}, \textbf{5.25x}, and \textbf{2.40x} compared to the GPU baseline.

\subsection{Latency Measurement}

We measure the latency of our accelerator by executing our FPGA kernel with a batch size of one. Figure~\ref{fig:evaluation} depicts latency measurements, design parameters for the accelerator are listed in Table~\ref{tab:params}. The red line indicates our application requirement of \SI{10}{\micro s}.

Our implementations reach the desired latency requirement of \SI{10}{\micro s}. We observe a small gap between end-to-end execution and compute results, which results from the AXI-4 interface overhead.  The large increase in latency for an event size of 128 points is largely caused by the reduced parallelism factor. 
In comparison to the GPU baseline, we achieve a substantial reduction in execution latency. Because the inference latency does not change between event sizes on the GPU, we assume the execution is bound by memory or orchestration overhead. Thus, a direct comparison would not be fair. 

\subsection{Resource Utilization}

\begin{figure}
   \small
   \centering
    \begin{tikzpicture}
        \pgfplotsset{
        every axis legend/.append style={
            at={(0.55,1.25)},
            anchor=north west,
            legend columns = -1}}
        \begin{axis}[
            width= \linewidth,
            height= 4cm,
            major x tick style = transparent,
            ybar=1pt,
            bar width=6pt,
            ymajorgrids = true,
            symbolic x coords={LUT,FF,DSP,BRAM,CLB},
            xtick = data,
            scaled y ticks = false,
            enlarge x limits=0.2,
            enlarge y limits=0.1,
            ymin=0,
            ymax=100,
            ytick={0, 25, 50, 75, 100},
            yticklabels={0~\%,25~\%,50~\%,75~\%,100~\%},
            minor y tick num=5,
            legend style={draw=none},
            legend image code/.code={%
      \draw[#1] (0cm,-0.1cm) rectangle (0.2cm,0.1cm);}]
            \addplot[style={white,fill=KITblue,mark=none}]
                coordinates {(FF, 11.91) (LUT,33.10) (CLB,45.28) (DSP,49.75) (BRAM,11.58)};
    
            \addplot[style={white,fill=KITorange,mark=none}]
                 coordinates {(FF, 20.34) (LUT,63.38) (CLB,84.35) (DSP,99.09) (BRAM,14.89)};
    
            \addplot[style={white,fill=KITcyan,mark=none}]
                 coordinates {(FF, 16.86) (LUT,44.66) (CLB,61.82) (DSP,49.75) (BRAM,16.86)};
    
            \addplot[style={white,fill=KITpurple,mark=none}]
                 coordinates {(FF, 28.64) (LUT,78.38) (CLB,97.33) (DSP,99.24) (BRAM,38.47)};

            \addplot[style={white,fill=KITbrown,mark=none}]
                coordinates {(FF, 35.56) (LUT,65.46) (CLB,95.02) (DSP,99.14) (BRAM,18.98)};
       
            \addlegendentry{A}
            \addlegendentry{B}
            \addlegendentry{C}
            \addlegendentry{D}
            \addlegendentry{E}

            \end{axis}
    \end{tikzpicture}
    \caption{Resource utilization for five firmware versions described in Table~\ref{tab:params}. Values are reported after P\&R.}
    \label{fig:evaluation:util}
\end{figure}
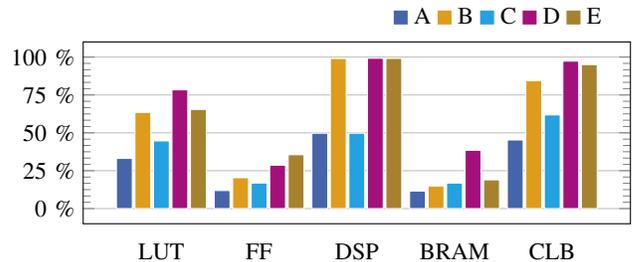

We report the system resource utilization of all five firmware versions in Figure~\ref{fig:evaluation:util}.
Notably, the utilization of Configurable Logic Blocks (CLBs) is much larger than both Flip-Flops (FFs) or Lookup-Tables (LUTs).
Generally, our model uses most of the available system resources.

\section{Discussion}
We have presented an approach to deploy graph-based PCNs on FPGAs, adhering to stringent real-time latency and throughput requirements. 
In this work, we developed dataflow actors for PCNs and implemented two graph processing elements for the operators (1) GraVNetConv and (2) condensation point clustering. 

A selected case study has been implemented on the AMD~VCK190 and has been evaluated on \textbf{hardware}.
In comparison to the GPU baseline, we achieve a relative throughput improvement in the range of \textbf{2.40x} to \textbf{5.25x}.
In addition, all our implementations reach the latency requirements of \SI{10}{\micro s}. 

To conclude, our work enables graph-based PCNs in real-time trigger applications for large scientific experiments. 

\ifthenelse{\equal{\blind}{false}}
{
\section*{Appendix}
The code of our FPGA implementation is available on Github, together with the respective firmware artifacts~\cite{pcnhlslib}.
}
\newpage

\printbibliography

\end{document}